# Electric-field-induced non-ergodic relaxor to ferroelectric transition in $BiFeO_3$-$x$$SrTiO_3$ ceramics


Leonardo Oliveira,[1] Jeppe Ormstrup,[1] Marta Majkut,[2] Maja Makarovic,[3,4] Tadej Rojac,[3,4] Julian Walker,[5] Hugh Simons[1, *]

[1]*Department of Physics, Technical University of Denmark, 2800 Kgs. Lyngby, Denmark*
[2]*ESRF – The European Synchrotron, Avenue des Martyrs, 38000 Grenoble, France*
[3]*Electronic Ceramics Department, Jožef Stefan Institute, 1000 Ljubljana, Slovenia*
[4]*Jožef Stefan International Postgraduate School, 1000 Ljubljana, Slovenia*
[5]*Department of Materials Science and Engineering, Norwegian University of Science and Technology, Trondheim, Norway*

Corresponding Author: leosdo@dtu.dk, husimo@fysik.dtu.dk



**Abstract**

While $BiFeO_3$-based solid solutions show great promise for applications in energy conversion and storage, realizing this promise necessitates understanding the structure-property relationship in particular pertaining to the relaxor-like characteristics often exhibited by solid solutions with polar-to-non-polar morphotropic phase boundaries. To this end, we investigated the role of the compositionally-driven relaxor state in (100-$x$)$BiFeO_3$-$x$$SrTiO_3$ [BFO-$x$STO], via *in situ* synchrotron X-ray diffraction under bipolar electric-field cycling. The electric-field induced changes to the crystal structure, phase fraction and domain textures were monitored via the $\{111\}_{pc}$, $\{200\}_{pc}$, and $½\{311\}_{pc}$ Bragg peaks. The dynamics of the intensities and positions of the $(111)$ and $(11\bar{1})$ reflections reveal an initial non-ergodic regime followed by long-range ferroelectric ordering after extended poling cycles. The increased degree of random multi-site occupation in BFO-42STO compared to BFO-35STO is correlated with an increase of the critical electric field needed to induce the non-ergodic-to-ferroelectric transition, and a decrease in the degree of domain reorientation. Although both compositions show an irreversible transition to a long-range ferroelectric state, our results suggest that the weaker ferroelectric response in BFO-42STO is related to an increase in ergodicity. This, in turn, serves to guide the development of BFO-based systems into promising platform for further property engineering towards specific capacitor applications.

**Introduction**

The increasing demand for energy necessitates more numerous and more efficient energy storage systems and devices. Furthermore, the range of conditions in which these devices must operate is rapidly broadening. To this end, oxide dielectric capacitors offer good thermal stability, excellent fatigue properties, and high energy densities with fast charge-discharge capabilities.[1,2] In particular, oxide perovskite-based relaxor ferroelectrics are a promising class of materials for high-k dielectric energy storage applications. These materials show unique polarization-electric field hysteresis behavior that allows both large field-induced polarization and breakdown strength while exhibiting low dielectric losses.[3] Although $PbTiO_3$-based solid-solutions stand as the benchmark for relaxor ferroelectrics, extensive research has focused on developing eco-friendly, lead-free materials offering competitive performance.[4,5] Significant progress has been made in recent years with $BaTiO_3$ (BTO), $(K_{1/2}Na_{1/2})NbO_3$ (KNN), $SrTiO_3$ (STO), $(Na_{1/2}Bi_{1/2})TiO_3$ (NBT), and $BiFeO_3$ (BFO) solid solution systems.[6] Among them, BFO-BTO offers the advantages of a high Curie temperature (*e.g.*, $T_c$ = 503 ºC for BFO-0.3BTO[7]), which in turn provides a wider temperature window for their eventual applications.[8]

One strategy for optimizing the energy storage properties of relaxor ferroelectrics consists of promoting compositionally-disordered regions and short-range correlations at the nanoscale by engineering a morphotropic phase boundary (MPB) between two different lattice symmetries.[9–11] Although BFO-based systems are known to possess polar-to-polar MPBs, such as in BFO-0.33BTO,[12] BFO-based systems may also exhibit polar-to-*non*-polar MPBs, such as in BFO-$x$STO.[13,14] Such polar-to-non-polar transitions are essential to engineering large energy densities via electric-field induced phase transitions from the non-polar to polar states, as similarly achieved with the electro-mechanical response in incipient piezoelectrics.[15] While such electric-field-induced transition are well understood in the context of antiferroelectric[16] and classical relaxor ferroelectric materials,[17] the polar-to-non-polar MPB in BFO presents a different structural character.[3,18] Understanding the unique balance between local structural disorder and long-range crystal symmetry is a significant challenge associated with BFO-based systems, and analogies with the more well understood PMN-PT materials have proven to be oversimplified.[19] Nonetheless, the complex structure-composition relationship is still key to successful property engineering.

In this work, we shed light on the structural mechanisms that underpin the relaxor-like behavior and electric-field-induced response of BFO-$x$STO solid solutions around the polar to non-polar morphotropic phase boundary, which lies at ~ 40% mol STO.[20] We carry out synchrotron X-ray diffraction experiments *in situ* during electric field cycling to understand the structural response to electric fields as a function of composition and, in particular, its transition from a ferroelectric character with macroscopic domains, which is essential for electromechanical applications, to the more relaxor-*like* behavior desired for high-density energy storage.

**Experimental methods**

The mechanochemical-assisted synthesis of BFO-35STO and BFO-42STO samples were conducted in a three-step process. The starting materials $Bi_2O_3$, $Fe_2O_3$, $TiO_2$, and $SrCO_3$ were pre-milled separately, then homogenized in a blend with equimolar ratios. Next, the powder mixture was subjected to high-energy planetary ball milling for 40 h. Finally, the activated powder was additionally homogenized in a planetary mill, pressed into pellets, and sintered at 1025 °C for 2 h with heating and cooling rates of 5 ºC/min. A detailed description about the synthesis method and its morphological features are given in Ref. 21.[21]

High-energy synchrotron X-ray diffraction experiments were then carried out at beamline ID11 of the European Synchrotron Radiation Facility (ESRF). The X-ray beam was rectangular, with dimensions 0.6 mm (horizontal) by 0.3 mm (vertical), and a wavelength of 0.31 Å (40 keV). Two-dimensional powder diffraction patterns were measured on a FReLoN CCD detector positioned 245 mm from the sample. The detector sampled a *q*-range of 4.24 Å$^{-1}$, spanning the first six diffraction rings, *i.e.* up to the $\{211\}_{pc}$. The samples were immersed in silicone oil within a custom-made sample holder, and a DC electric field was applied *in situ* in the following scheme: starting from 0 kV/mm, a triangular bipolar cycle from -14 kV/mm to 14 kV/mm ($E_{max}$) over a time (*t*) of 1620 s (27 minutes). Images were acquired every two seconds. This voltage ramping scheme was required to minimize the likelihood of dielectric breakdown during the application of the electric field. The raw 2D images were corrected for detector distortions and background intensity, and azimuthally segmented into 36, 10°-wide bins.[22]

The analysis addressed the electric-field and orientation dependent changes to the $\{111\}_{pc}$ and $\{200\}_{pc}$ Bragg peaks. The $\{111\}_{pc}$ peak corresponds to the direction of spontaneous polarization in the rhombohedral polar phase of BFO and its derivatives, whereas the $\{200\}_{pc}$ peak is associated with intergranular elastic strain coupling.[22] Individual peak profiles were fit with pseudo-Voigt profiles using least-squares minimization, implemented in Matlab. Prior to the application of the *E*-field, we note that the $\{111\}_{pc}$ of BFO-35STO required two pseudo-Voigt profiles to accurately reproduce the peak shape, while only a single profile was needed for BFO-42STO. Above a certain critical electric field, two profiles were needed for both compositions.

The polarization, current and strain versus electric field hysteresis loops (*P-E, J-E* and *S-E*, respectively) were measured using an aixACCT TF 2000 analyzer (aixaCCT Systems GmbH, Aachen, Germany) equipped with a laser interferometer (Type SP-S, SIOS Meßtechnik GmbH, Ilmenau, Germany). The samples were immersed in silicone oil and the hysteresis loops were recorded by applying to the samples three single sinusoidal waveforms of 10 Hz at selected electric field amplitudes with a delay between the cycles with zero field of a few seconds.
The *P-E* loops shown in this paper are those obtained during the second field cycle.

**Results**

**Figure 1(a, b)** shows the XRD patterns measured prior to the application of the electric field, both of which exhibit characteristics consistent with a pseudo-cubic average symmetry. At the same time, there is an overall shift of the diffraction peaks toward higher $2\theta$ associated with the increased fraction of STO (*e.g.* for $\{111\}_{pc}$, the $\Delta 2\theta$ is 0.02°), which indicates a small reduction in unit cell volume. Atomic coordinates do not show significant shifts from the fixed cubic Wyckoff sites, and as such the fitting of the $\{111\}_{pc}$ peak in this region gives satisfactory residuals using either one or two peak profile functions. However, both compositions show evidence of a weak satellite reflection corresponding to the $\frac{1}{2}\{311\}_{pc}$ lattice planes, which originates from out-of-phase $a^-a^-a^-$ octahedral tilting and suggests the presence of a phase isostructural with the room-temperature BiFeO$_3$ phase with *R*3*c* symmetry.[23–25] The presence of this satellite peak in the absence of pronounced $\{111\}_{pc}$ peak splitting suggests short-range order with an *R*3*c* local average symmetry. Similar effects are regularly observed in well-characterized relaxor ferroelectrics, such as NBT-BT.[26] Although none of the fundamental reflections exhibit significant or orientation-dependent splitting in this initial state, **Figure 1(c, d)** highlights that the $\frac{1}{2}\{311\}_{pc}$ satellite associated with the *R*3*c* symmetry exhibits

orientation-dependent residual strain, *i.e.* texture, which is particularly significant for azimuthal angles ($\psi$) higher than 30º.

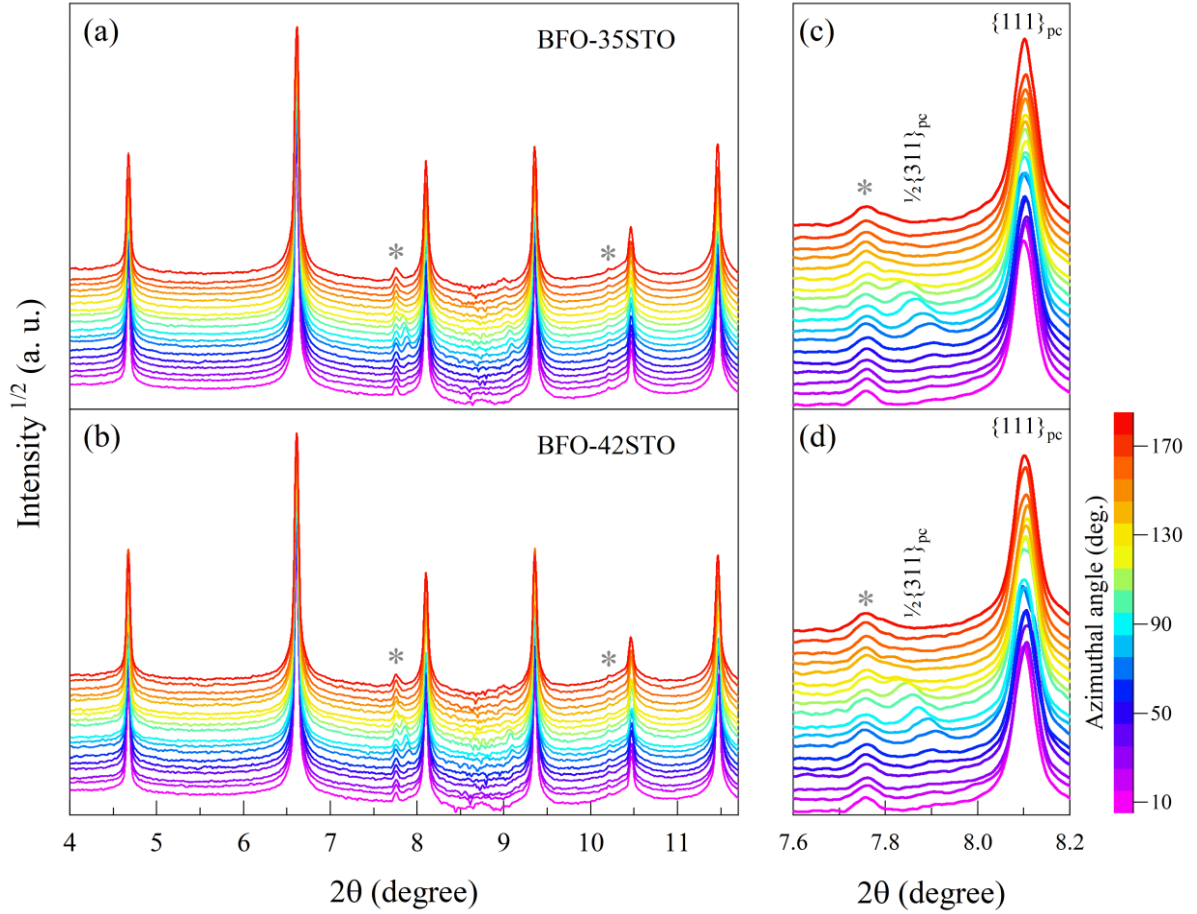

**Figure 1.** Azimuthal angle-dependency of the XRD patterns for (a) BFO-35STO, and (b) BFO-42STO. The peaks marked with the star symbol (*) belong to a minor secondary phase, most likely Sillenite ($Bi_{25}FeO_{40}$). The panels (c) and (d) highlights the $\{111\}_{pc}$ and the ½$\{311\}_{pc}$ reflections for BFO-35STO and BFO-42STO, respectively.

To understand the evolution of this structure during continued electric field cycling, we next acquired XRD patterns *in-situ* during the application of bipolar electric fields to the samples along the polarization direction. In **Figure 2(a)**, the contour map of the $\{111\}_{pc}$ reflections for BFO-35STO sample under bipolar field cycling at $\psi = 0º$ reveals an apparent shift toward lower $2\theta$ with increasing field magnitude. Upon closer inspection of the $\{111\}_{pc}$ intensity profiles (**Figure 2(b)**), we also notice an asymmetry that shifts from the low-$2\theta$ side in the virgin state, to the high-$2\theta$ side after the first application of large field magnitudes and for all subsequent bipolar cycles. Deconvoluting the $\{111\}_{pc}$ intensity profiles via fitting reveals that both this apparent variation in $2\theta$ and asymmetry are not the result of a unit-cell expansion, but a ferroelastic reorientation of the non-180º domains walls. This is evidenced in **Figure 2(b)** from the fitting results of selected $\{111\}_{pc}$ reflections at different electric field magnitudes, which confirm the corresponding intensity interchange of the (111) and (11$\bar{1}$) peaks.

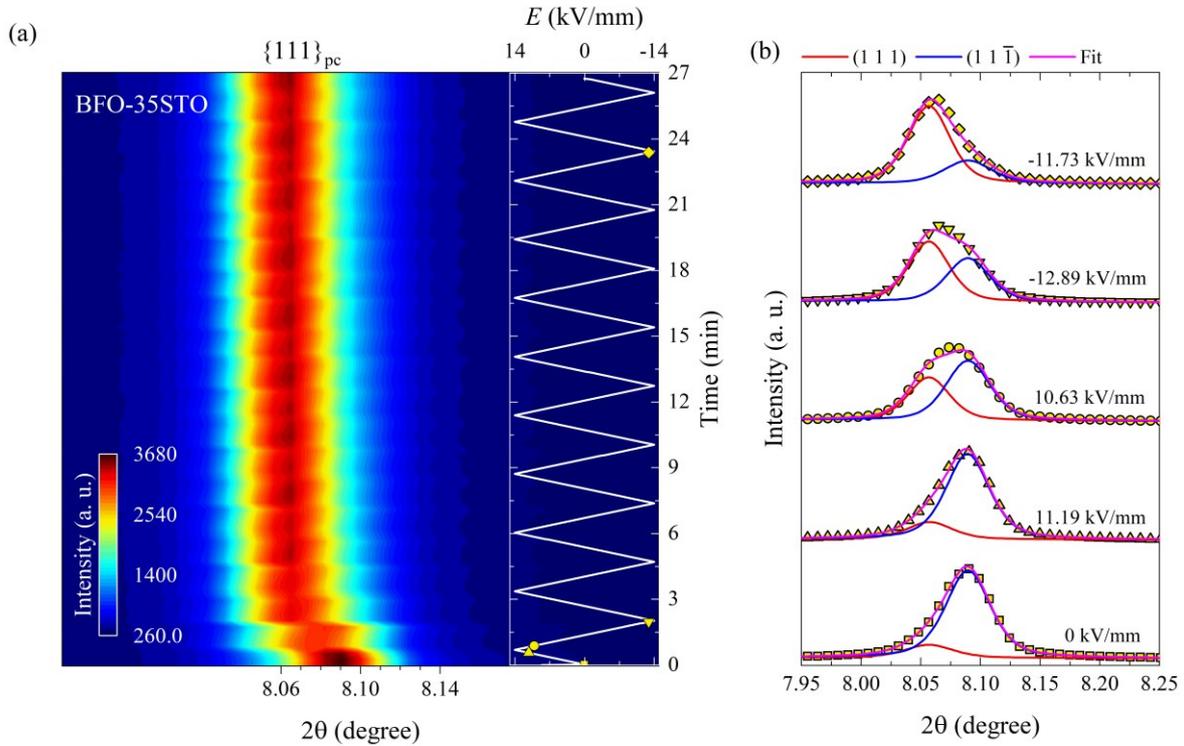

**Figure 2.** Bipolar *E*-field cycling along the polarization direction for BFO-35STO. (a) Contour map of the $\{111\}_{pc}$ reflections at $\psi = 0°$. (b) Deconvoluted peak fitting evolution for selected *in-situ* diffracted intensity distributions. The insert in (a) represents the $E(t)$ evolution, where the marked yellow symbols correspond to the fitted intensity distributions in (b).

**Figure 3** shows the relative position and intensity interchange of the (111) and (11$\bar{1}$) peaks in BFO-35STO during electric field cycling, and reveals that the extended bipolar cycling is characterized by three regions over time: (1) an initial region without any evidence of non-180° ferroelectric switching, followed by (2) a region in which an *E*-field induced transformation to a ferroelectric state takes place, and (3) a region were the material is predominantly ferroelectric, and the structural changes are dominated by ferroelastic switching. In the first (green) region, which is associated with the virgin state until the critical field ($E_{CR}$), the applied *E*-field does not induce any significant structural changes, such as variations in the integrated intensity ($I_{int.}$) or shifts in $2\theta$ position of the split $\{111\}_{pc}$ peaks. The second (orange) region, extending from $E_{CR}$ to approximately the minimum negative field, is highlighted by a significant decrease in $I_{int\,(11\bar{1})}$, which denotes the development of an *E*-field induced texture.

Finally, in the third (blue) region, which completes the first cycle, the system begins to show signatures of ferroelastic back-switching, as would be expected in a canonical ferroelastic-ferroelectric switching. The complete intensity interchange is reached after $E = -11.2$ kV/mm. Then, the classic long-range ferroelectric/ferroelastic behavior takes place after the first $E$-field minima, maintaining a non-random domain orientation distribution for the successive poling cycles.

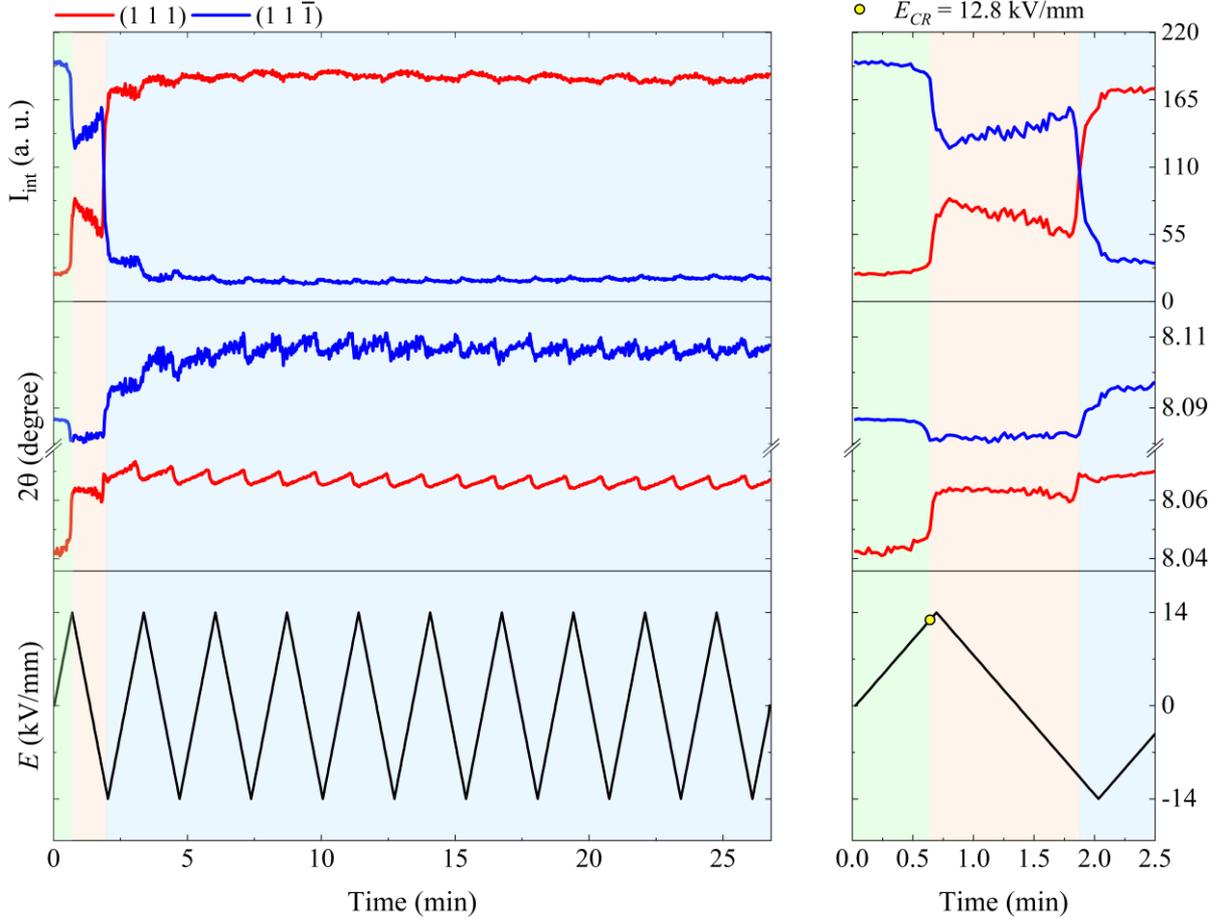

**Figure 3.** Evolution of the fitted $I_{int}$ and $2\theta$ over $E(t)$ cycling for BFO-35STO. The right panel is an enlarged view of the first $E$-field cycle (from 0 to 2.5 min), while the right panel covers the entire time scale (from 0 to 27 min). The green area marks the region where signatures of ferroelectricity are absent, i.e. without $\{111\}_{pc}$ intensity interchange, while the orange and blue area denote regions where the onset of an electric-field-induced phase transformation and ferroelastic domain interchange take place, respectively.

In contrast to BFO-35STO, the $E$-field dependent diffracted intensity distributions of the $\{111\}_{pc}$ reflections at $\psi = 0°$ for BFO-42STO reveal a more disordered average pseudo-cubic structure, as well as a lower degree of $E$-field induced ferroelastic reorientation after extended bipolar cycling. These effects are illustrated in **Figure 4**, where the $\{111\}_{pc}$ reflection from 0 kV/mm to 14 kV/mm (*i.e.* before the first $E_{max}$) remains a single peak, *i.e.* without any evidence of the low-$2\theta$ asymmetry associated with the (111) peak. As a result, we refer to this peak as $(111)_{pc}$ to distinguish the lack of splitting. As the *E*-field decreases after the first $E_{max}$, the $(111)_{pc}$ peak begins to split into the $\{111\}_{pc}$ family of peaks, *i.e.* (111) and $(11\bar{1})$ doublets, as the pseudo-cubic symmetry is gradually broken and the material transforms into the polar *R3c* phase. At the same time, this splitting is associated with a discontinuous increase of unit-cell volume, indicated by the shift of the peak maxima position towards lower $2\theta$ (also seen by the abrupt shift of (111) peak position in **Figure 4(a)**), which is consistent with a *E*-field induced first-order paraelectric to ferroelectric phase transition.[27] However, the deconvoluted peaks at $E = -10.1$ kV/mm after multiple *E*-field cycles show that, although there is a pronounced orientation favoring the (111) orientation parallel to the *E*-field, the degree of this preferred orientation is not as high as for BFO-35STO, given that the $(11\bar{1})$ peak has intensity comparable to the (111) (see fitted profiles on **Figure 3(b)** and **Figure 4(b)**). In other words, increasing the fraction of STO reduces the ability to achieve a saturated polarization under applied *E*-fields.

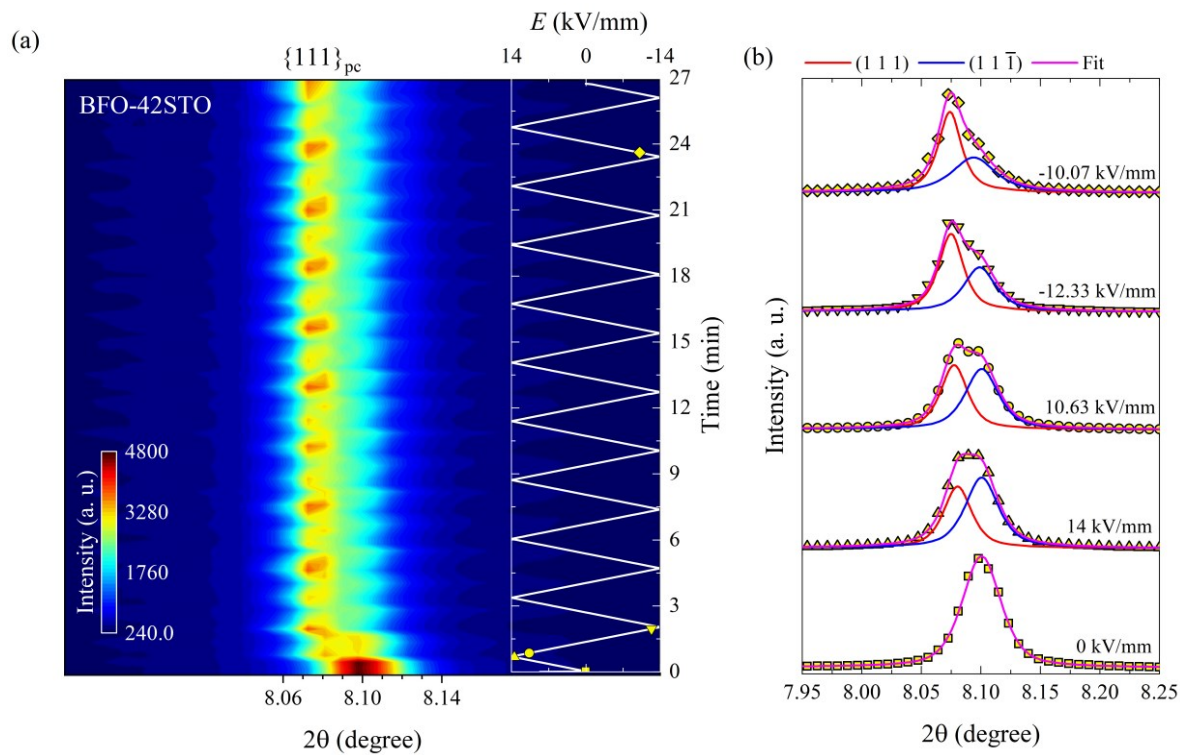

**Figure 4.** Bipolar *E*-field cycling along the polarization direction for BFO-42STO. (a) Contour map of the $\{111\}_{pc}$ reflections at $\psi = 0°$. (b) Deconvoluted peak fitting evolution for selected *in-situ* diffracted intensity distributions. The insert in (a) represents to the $E(t)$ evolution, being the marked yellow symbols correspondent to the fitted

intensity distributions in (b). In contrast to BFO-35STO, between 0 and 14 kV/mm the $\{111\}_{pc}$ peak is not visibly split before the first $E_{max}$.

The $(111)_{pc}$ peak parameters for BFO-42STO before the splitting and the $\{111\}_{pc}$ peak fitting parameters after the splitting are shown in **Figure 5**. The evolution of these parameters indicates the structure is initially non-ferroelectric, before transforming to ferroelectric with further *E*-field cycling. However, unlike in BFO-35STO, there is no clear interchange of intensity between the $\{111\}_{pc}$ reflections, because the (111) intensity is most of the time greater than that of the $(11\bar{1})$. Therefore, although the time required for the $I_{int}$ oscillations to stabilize around an approximate constant mean value is lower in BFO-42STO (1.7 min) than in BFO-35STO (3.9 min), the resultant dipole realignment in BFO-42STO is poorly affected by the *E*-field application, given that the ratio of the deconvoluted $I_{int}$ is lower. This, in turn, implies a heterogeneous mesoscopic domain structure with a lower degree of *E*-field induced texture.

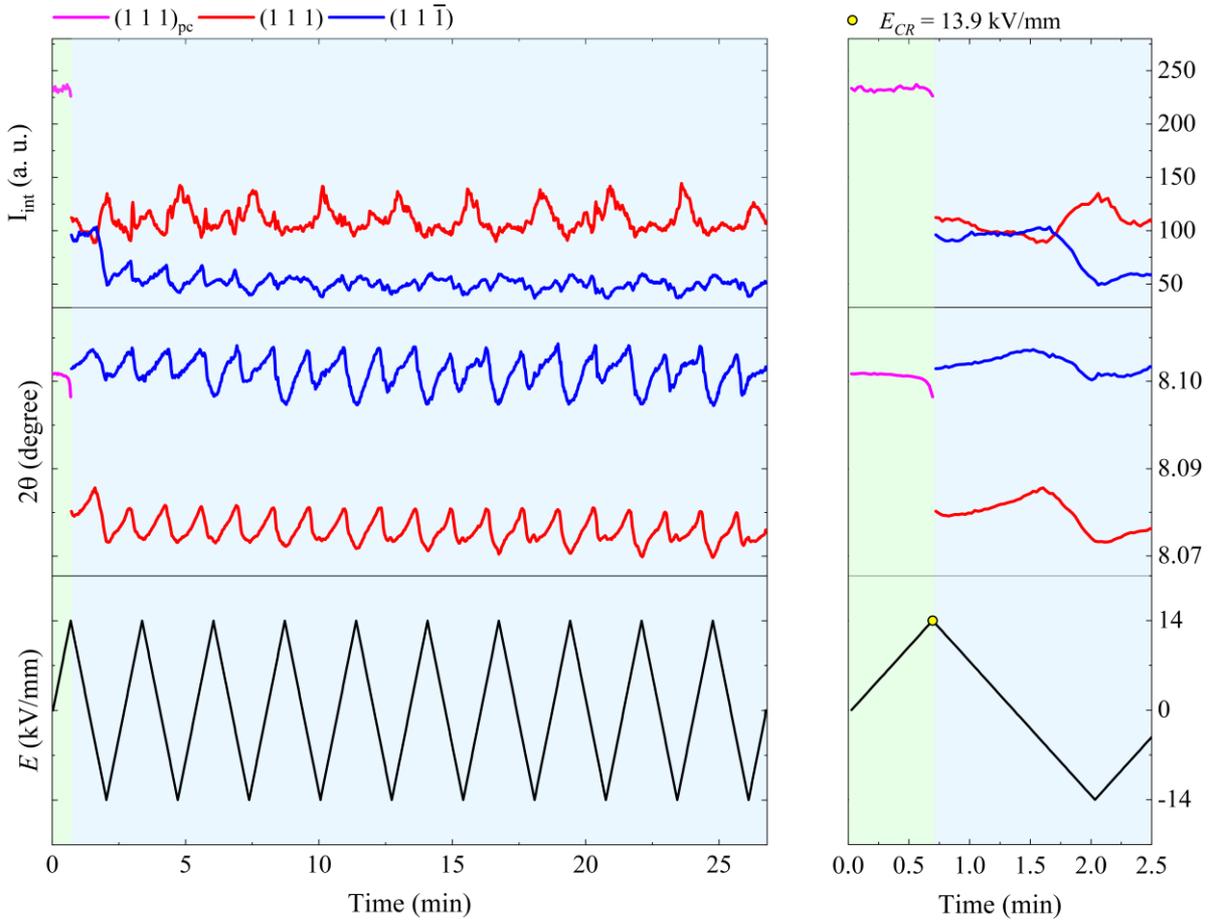

**Figure 5.** Evolution of the fitted $I_{int}$ and $2\theta$ over $E(t)$ cycling for BFO-42STO. The right panel is an enlarged view of the first *E*-field cycling (from 0 to 2.5 min), while the right panel covers the entire time scale (from 0 to 27

min). The green area denotes a region where the ferroelectric behavior is absent, implying the $I_{int}$ related to an unsplit $(111)_{pc}$ peak. The blue area denotes a region where $(111)_{pc}$ splits into two peaks, denoting the appearance of the rhombohedral distortion, and ferroelastic domain interchange by the periodic $I_{int}$ oscillation.

In order to correlate the microstructural changes with the macroscopic properties, the relationship between the macroscopic *P-E*, *J-E*, and *S-E* hysteresis loops, lattice strain (*ε*), and texture are compared in **Figure 6**.

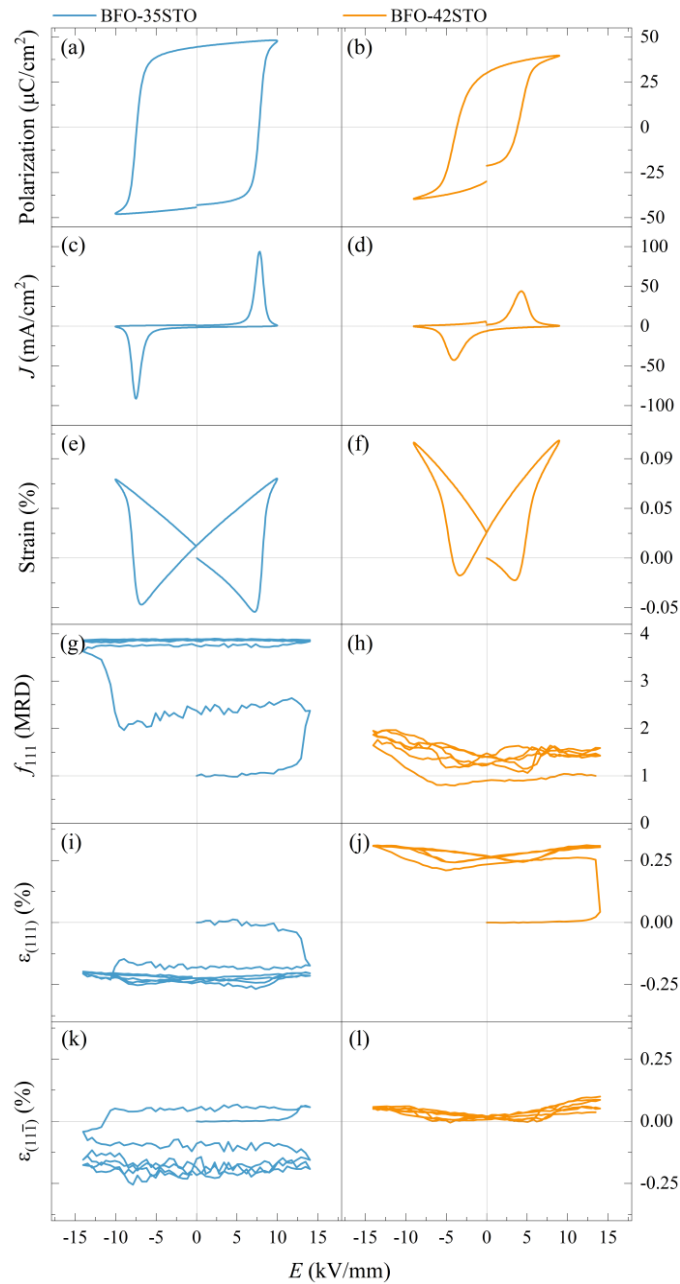

**Figure 6.** Macroscopic *P-E* (a, b), *J-E* (c, d), and *S-E* (e, f) hysteresis loops, $f_{111}$ texture indexes (g, h), and lattice strain for (111) (i, j), and $(11\bar{1})$ domains (k, l), for BFO-35STO and BFO-42STO, respectively. The cycles shown in (g-l) corresponds to the first two bipolar periods and is representative of the entire dataset.

As shown in **Figure 6(a-f)**, typical ferroelectric *P-E, J-E,* and *S-E* responses are observed in both the BFO-35STO and BFO-42STO samples. However, increasing the STO fraction reduces both the maximum ($P_{max}$) and remanent ($P_r$) polarization, as well as the coercive field ($E_c$). Concomitantly, the decrease and widening of the *J-E* peak in BFO-42STO is associated to a weakening of the induced ferroelectric ordering as the domains are less prone to switch. In the *S-E* loops, the effect of adding STO is observed by the reduced negative strain in BFO-42STO, while the peak-to-peak strains retain similar values for both BFO-35STO and BFO-42STO. This is consistent with a shift from a ferroelectric/ferroelastic switching behavior toward a strain behavior more characteristic for relaxor ferroelectrics.[28]

The *in situ* diffraction data contains quantitative microstructural information regarding the changes in the ferroelastic reorientation, which may be analyzed via changes in the integrated intensity ratios during poling. In this way, domains aligned to the *E*-field will increase in volume and, consequently increase the corresponding XRD peak intensity. Therefore, the extrinsic strain contribution is a direct consequence of domain-wall motion.[29,30] The texture evolution in the non-180º ferroelectric domain walls was evaluated by the multiple random distribution (MRD) model through the $f_{111}$ coefficient for a rhombohedral system by **Eq.1**

$$f_{111}(MRD) = 4 \times \frac{I^E_{int(111)}/I^0_{int(111)}}{I^E_{int(111)}/I^0_{int(111)} + 3 \times I^E_{int(11\bar{1})}/I^0_{int(11\bar{1})}} \qquad (1)$$

where $I^0_{int(111)}$ and $I^E_{int(11\bar{1})}$ are the integrated intensity at the virgin state and under electric field, respectively. The MRD index ranges from 0 to 4, where $f_{111} = 1$ represents the unpoled state with randomly orientated domains while $f_{111} = 4$ represents the poled state with all the domains parallels to the *E*-field, with 4 being the number of non-parallel polarization directions allowed by the rhombohedral symmetry of the unit cell.[22,31]

The evolution of the $f_{111}$ index illustrates the *E*-field induced texture influenced by the STO content. **Figure 6(g)** shows the texture evolving from a random domain distribution (at the virgin state) to a more poled state just after the initial cycle is finished. This slower and irreversible variation of the ferroelastic reorientation (and after on the field-induced lattice strain) is similar to the "wake-up" behavior observed in Sm-doped BFO, where the slow polarization and strain development are attributed to an *E*-field induced phase transformation and point defect redistribution.[32] After the entire poling cycle, the texture is close to the saturation, displaying $f_{111_{max}} \approx 3.89$. In contrast, given that BFO-42STO shows an average pseudo-cubic structure from the virgin state to $E_{CR}$, the $f_{111}$ index cannot be meaningfully determined within this first interval. Additionally, the and the critical domain interchange field are not evident for this composition. Consequently, **Figure 6(h)** shows that the texture evolution is limited, demonstrating a reduced domain population just after $E_{CR}$, with $f_{111_{min}} \approx 0.79$, before increasing to $f_{111_{max}} \approx 2.07$ despite an asymmetric cyclic variation toward negative electric field values.

At the same time, the *in situ* diffraction data also contains quantitative information regarding the elastic strain and piezoelectric effect, which may be analyzed from the shifts in peak positions during poling. The intrinsic field-induced lattice strain is obtained by **Eq.2**

$$\varepsilon_{hkl} = \frac{d^E_{hkl} - d^0_{hkl}}{d^0_{hkl}} \quad (2)$$

where $d^E_{hkl}$ and $d^0_{hkl}$ are the interplanar spacings during the *E*-field application and in the virgin state, respectively, determined from the $2\theta$ shifts via Bragg's law (**Figure 3** and **Figure 5**).

As expected, the butterfly loops in **Figure 6(i-l)** for $\varepsilon_{\{111\}_{pc}}$ confirms that both compositions have a clear intrinsic piezoelectric response. **Figure 6(i, k)** shows the *E*-field-induced intrinsic strain (both $\varepsilon_{(111)}$ and $\varepsilon_{(11\bar{1})}$) is predominantly negative for BFO-35STO, which also correlates with the $\{111\}_{pc}$ intensity interchange denoted by the $f_{111}$ index (compare with **Figure 6(g)**). However, this induced negative strain is not observed in BFO-42STO, where instead the $\varepsilon_{(111)}$ response initially experiences a large positive strain before the hysteric behavior while the $\varepsilon_{(11\bar{1})}$ oscillates around 0.

This negative strain behavior in BFO-35STO is likely related to a complex intergranular microstrain effect during the wake-up regime of the first few cycles. In order to associate the strain induced by the switching of non-180° domain walls under the applied *E*-field with intergranular elastic strain coupling, the evolution of the $\{200\}_{pc}$ diffraction peaks are evaluated for both compositions.[33] **Figure 7(a, b)** shows the time and *E*-field dependent variation of the $\{200\}_{pc}$ peak where, after the first $E_{max}$, the position of the peaks is shifted towards lower angles in comparison with the initial state, indicating the strains are mainly tensile (positive) with respect to the *E*-field direction. In **Figure 7(c, d)**, the $\{200\}_{pc}$ strain hysteresis shows that both compositions exhibit tensile strains during initial *E*-field application, with BFO-35STO exhibiting a slightly larger initial lattice strain compared to BFO-42STO. A comprehensive illustration of the *E*-field- and *t*- dependent strain is shown in **Figure S1.** The overall $\varepsilon(E,t)$ evolution for BFO-35STO reveals that, after the wake-up-like process is complete (in which the polarization develops over the first few cycles), $\varepsilon_{\{200\}_{pc}} > \varepsilon_{(11\bar{1})} \approx \varepsilon_{(111)}$. For example, at 10 min (-12.9 kV/mm) the absolute strain values were 0.16% and -0.20% for $\{200\}_{pc}$ and $(111/11\bar{1})$, respectively (see the dashed lines in **Figure S1**). As expected, this implies that the strain anisotropy imposes a large tensile strain on the $\{200\}_{pc}$ grains to minimize the compressive strain along the polar [111] direction, and indicates that substantial intergranular stresses are developed during the poling process.[34,35] In BFO-42STO, however, these induced tensile strain in the $\{200\}_{pc}$ oriented grains are not nearly as significant, with $\varepsilon_{(111)} > \varepsilon_{\{200\}_{pc}} > \varepsilon_{(11\bar{1})}$ after the initial poling cycles. For example, at 14 min (12.9 kV/mm) the strain values are 0.32%, 0.14%, and 0.05% for (111), $\{200\}_{pc}$, and $(11\bar{1})$, respectively (see the dashed lines in **Figure S1**). Ultimately, we observe that the intergranular strains during wake-up are smaller in magnitude for BFO-42STO than for BFO-35STO. This reduced strain is correlated with a reduced degree of poling (visible in the $f_{111}$ trace in **Figure 6(g, h)**) in BFO-42STO, and its correspondingly increased ergodicity.

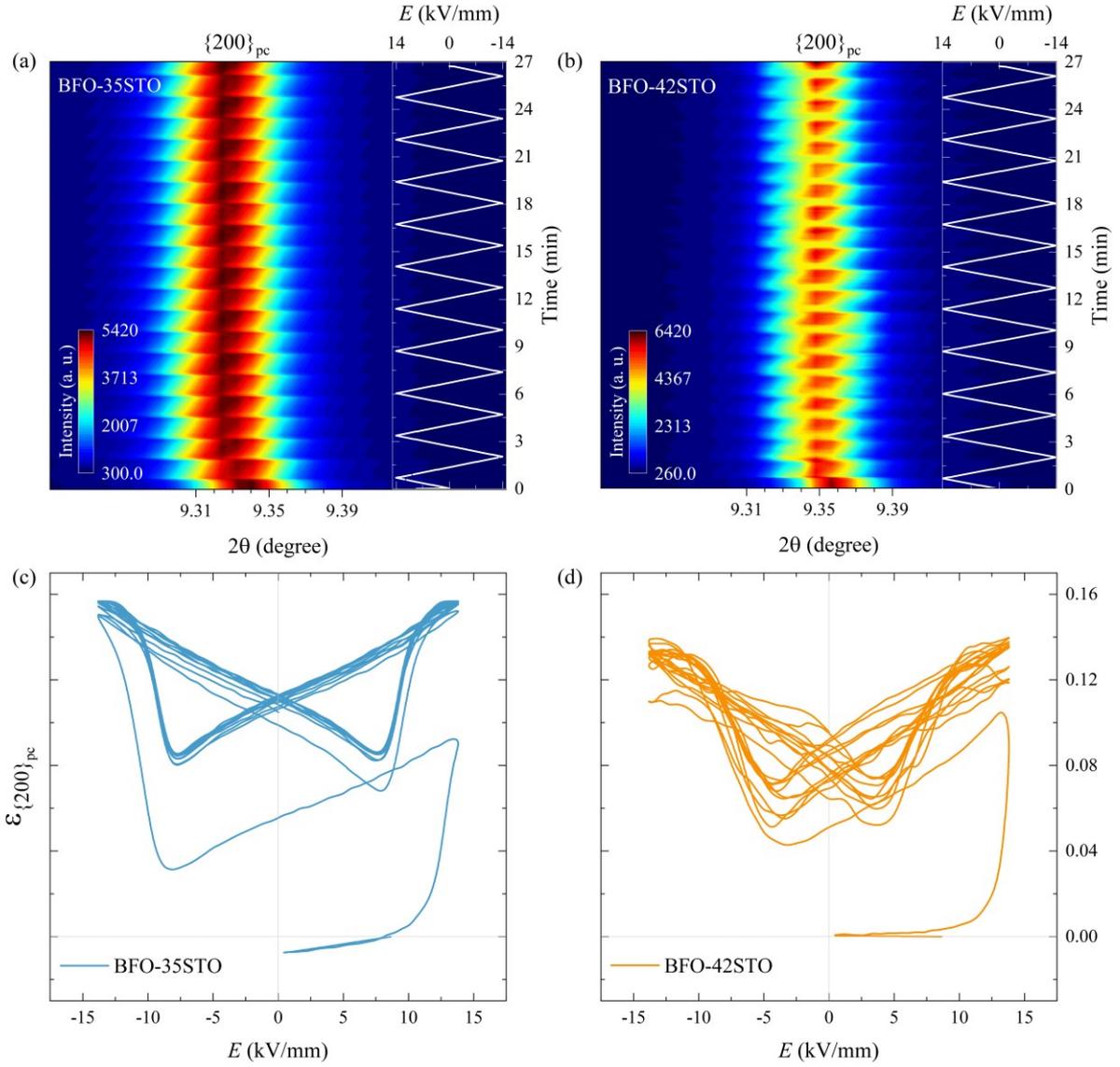

**Figure 7**. Contour map of the $\{200\}_{pc}$ reflections at $\psi = 0°$ under bipolar $E$-field cycling for (a) BFO-35STO, and (b) BFO-42STO. Lattice strain hysteresis loop for (c) BFO-35STO, and (d) BFO-42STO. In both compositions the $\{200\}_{pc}$ peak remains unsplit during the entire field cycling.

**Discussion**

Adding heterovalent $Sr^{2+}$ and $Ti^{4+}$ ions in the $BiFeO_3$ structure induces both chemical and charge disorder on the Bi and Fe lattice sites, respectively. Depending on the synthesis method, several reports have shown that the crystal structure of BFO-$x$STO for $20 < x < 60$ mol % tends to be pseudo-cubic.[36,37] However, discussions regarding the structural evolution, its correlation with the material chemistry and its macroscopic response remains scarce. One of the main reasons for this is because the reliable processing of BFO-based ceramics with low electrical conductivity, dielectric loss, leakage current, and high breakdown field remains challenging.[37] In this context, it was recently shown that minimizing intrinsic defects such as $Fe^{4+}$ and oxygen vacancies are beneficial to achieving poled ferroelectric states with saturated polarization.[38]

The diffraction patterns and macroscopic property measurements indicate that increasing the fraction of STO makes the structural evolution of the material more consistent with a relaxor ferroelectric. Both compositions exhibit XRD patterns consistent with a pseudo-cubic average structure, and there is a slight reduction in unit cell volume in BFO-35STO compared to BFO-42STO. This is expected with the substitution of larger $Fe^{3+}$ for smaller $Ti^{4+}$ ions and drives the structure to be more cubic-*like*. But, in spite of this cubic-*like* average structure, the diffraction measurements from both compositions exhibit weak, half-order superlattice reflections corresponding to the polar *R*3*c* phase expected for pure $BiFeO_3$. While it is common to explain such superlattice reflections in the presence of pseudo-cubic average structures in terms of Fe-rich regions with *R*3*c* symmetry within a Ti-rich matrix with pseudo-cubic symmetry (*i.e.* regions with smaller and larger unit cell volume, respectively), such compositional heterogeneity does not occur in such mechanochemically-prepared samples.[21]. This then leaves the alternative explanation: that the *R*3*c* structure is locally disordered (*e.g.* via the presence of nanodomains) but with an average cubic symmetry.

The possibility of a strongly disordered structure reducing the maximum and remanent polarization and coercive field of a ferroelectric (as also observed in $BiFeO_3$-$BaTiO_3$ solid solutions[39]) agrees with our previous study,[21] in which we observed: i) a broad diffuse phase transition, ii) frequency dispersion peak in $\varepsilon'(T)$, iii) $T_m$ shifting to lower temperatures with higher STO concentrations, iv) sharp decrease in remanent polarization and $d_{33}$ coefficient for $35 < x < 42$. Such a behavior indicates frustration of the long-range polar order, resulting in a relaxor-*like* ferroelectric state.[40] In addition, the $Sr^{2+}$ is also in principle expected to decrease the *A*-site off-center displacement of the ions trapped within oxygen dodecahedral cages, resulting in weak hybridization between $Bi^{3+}$ $6s^2$ lone pairs and O $2p$ orbitals and a reduction of the covalent binding character.[41]

At sufficiently high *E*-fields above the coercive-field, the non-ferroelectric regime vanishes, yielding an irreversible polar order in both samples. Interestingly, several differences can be seen in the structural dynamics by increasing the amount of STO. We observed that the $E_{CR}$ required to promote the ferroelectric transition in BFO-42STO ceramic is 1.1 kV/mm higher (*i.e.* the limit of the green regions in **Figure 3** and **Figure 5**). In this context, it may imply the site occupation in the ABO$_3$ structure of BFO-35STO is more ordered, favoring the dynamics of the local dipoles have more freedom to preferentially re-orientate (extension/rotation) along the <111> direction, indicative of a flattened free-energy profile in this composition.[9] Therefore, the disrupted period without signatures of ferroelectric switching is lower (0.64 min for BFO-35STO vs. 0.7 min for BFO-42STO), which allows larger Fe/Ti and Bi/Sr displacements from their central positions. Although distinct models are still under debate to correlate the origin and mechanisms involving polar nanoregions,[19] our results indicates that the presence of relaxor phase in BFO-$x$STO ceramics emerges due to the increased charge disorder.[42] Interestingly, recent *in situ* poling studies in BFO-40STO with $R3c$ cores surrounded by clamped pseudo-cubic shells have not demonstrated significant variations until 10 kV/mm.[43] Consequently, in that study, the field-induced distortions from short-to-long range could not be investigated in detail, and instead revealed only electrostrictive behavior.

Our results show that the applied field promotes an irreversible crossover from a non-ferroelectric, relaxor-*like* behavior to an ordered long-range polar order. With further cycling of the electric field, we also observe a distinct evolution occurring at $E$-fields above that of the domain interchange. In BFO-35STO, the domains are strongly sensitive to $E$-fields. However, with further destabilization of the long-range $R3c$ ordering in the BFO-42STO sample, the relaxor properties are enhanced, and preferential growth of ferroelectric domains are less prone to nucleate due to increased ergodicity. This effect is mainly observed in the $f_{111}$ and $\varepsilon_{(11\bar{1})}$ profile, where the wake-up behavior vanishes because the 71º and 109º ferroelastic domains become less easily switched and, consequently, reduce the degree of field-induced texture.[22,44] This feature is consistent with the previously measured permittivity behavior (see **Figure S2**), where the diffuse frequency-dependent peak shift towards lower temperatures as the ergodicity increases, i.e., with increasing concentration of STO in BFO-STO.[21]

**Conclusions**

In summary, the crossover from relaxor to long-range polar state in BFO-$x$STO solid-solutions has been investigated by *in situ* synchrotron XRD. Distinct features of the $\{111\}_{pc}$ reflections evolve as the chemical disorder increases, in particular pertaining to the destabilization of the average pseudo-cubic structure during $E$-field cycling. With increasing STO content, higher $E$-fields are required to overcome the non-ergodic state, presumably due to random multi-site distribution of cation species and a more pronounced ergodic relaxor state.

In addition, while the dominant feature in the BFO-35STO is a well-defined ferroelastic reorientation, in BFO-42STO the ferroelectric ordering is weakened due to the increase in charge disorder and the subsequent relaxor behavior it induces. Consequently, the preferential growth of field-induced ferroelectric macrodomains is not prone to take place, inhibiting polarization enhancement along <111> as the ergodicity increases. Therefore, in BFO-42STO the ergodicity limits the development of a wake-up behavior. Ultimately this influence of STO in inducing a polar-non-polar MPB while simultaneously tuning the ergodicity of the polar response is expected to open new, targeted approaches for engineering energy storage materials, in particular for applications requiring broad temperature regimes.

**Conflicts of interest**

The authors declare no conflicts of to declare.

**Acknowledgements**

The authors are grateful to ESRF for providing beamtime through the user access programme. L.O. and H.S. acknowledge financial support from ERC Starting Grant #804665. J.W. acknowledge the Walker Science for Life foundation. T.R. would like to thank the Slovenian Research Agency for financial support (research core funding P2-0105 and research project J2-3042).

# Supporting information

**Electric-field-induced non-ergodic relaxor to ferroelectric transition in BiFeO$_3$-$x$SrTiO$_3$ ceramics**

Leonardo Oliveira,[1] Jeppe Ormstrup,[1] Marta Majkut,[2] Maja Makarovic,[3,4] Tadej Rojac,[3,4] Julian Walker,[5] Hugh Simons[1, *]

[1]Department of Physics, Technical University of Denmark, 2800 Kgs. Lyngby, Denmark
[2]ESRF – The European Synchrotron, Avenue des Martyrs, 38000 Grenoble, France
[3]Electronic Ceramics Department, Jožef Stefan Institute, 1000 Ljubljana, Slovenia
[4]Jožef Stefan International Postgraduate School, 1000 Ljubljana, Slovenia
[5]Department of Materials Science and Engineering, Norwegian University of Science and Technology, Trondheim, Norway

Corresponding Author: leosdo@dtu.dk, husimo@fysik.dtu.dk

**Content**

**$\varepsilon(E, t)$ dependency**

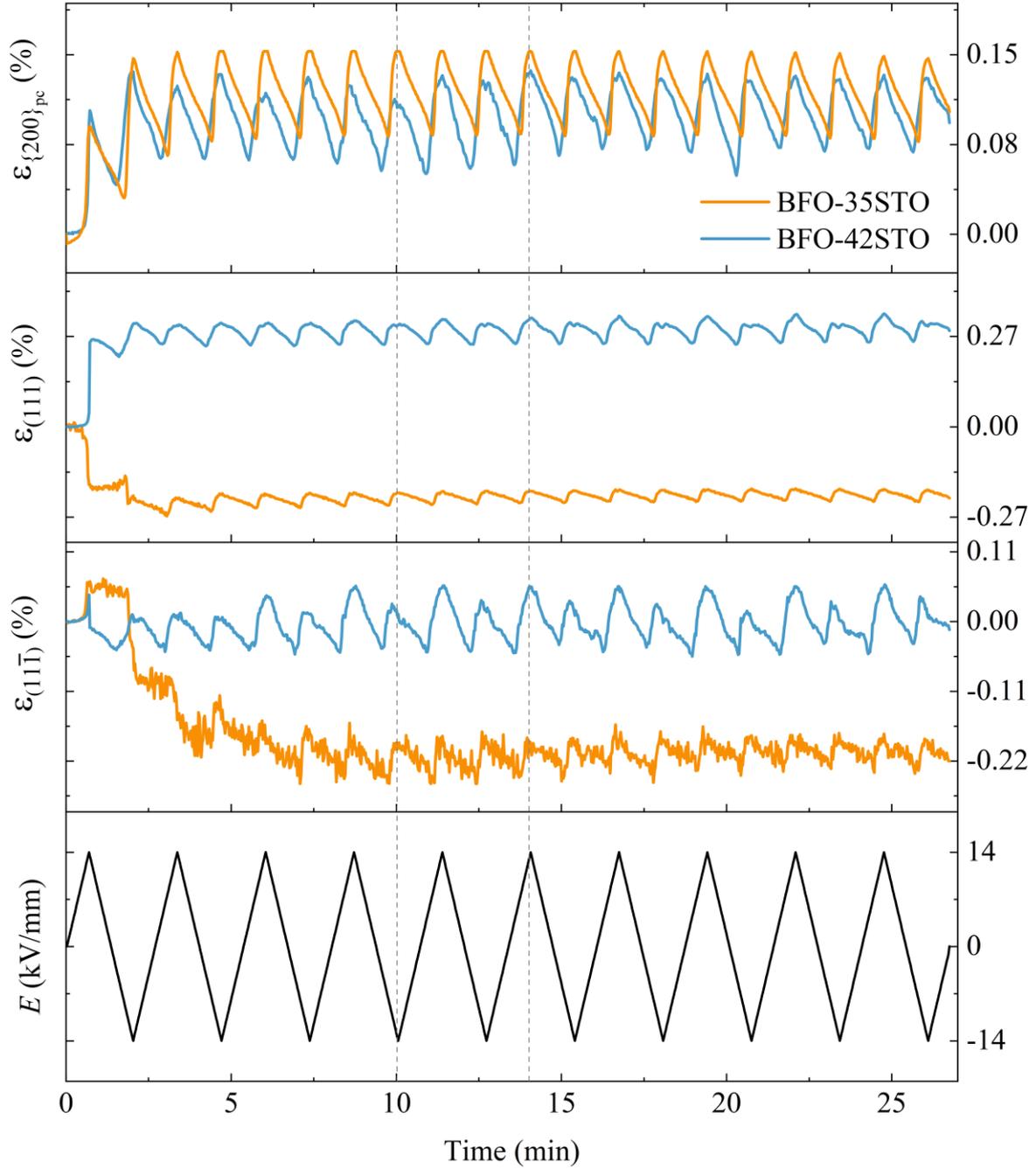

**Figure S1**. Comprehensive time and $E$-field evolution of $\varepsilon_{\{111\}_{pc}}$ and $\varepsilon_{\{200\}_{pc}}$ for the entire field cycling. The $\varepsilon(E, t)$ dependency shows a general behavior for times higher than the initial wake-up cycle. This trend is exemplified from the dashed lines at 10 min and 14 min. One may conclude that $\varepsilon_{\{200\}_{pc}} > \varepsilon_{(11\bar{1})} \approx \varepsilon_{(111)}$ for BFO-35STO, while $\varepsilon_{(111)} > \varepsilon_{\{200\}_{pc}} > \varepsilon_{(11\bar{1})}$ for BFO-42STO.

**Dielectric permittivity**

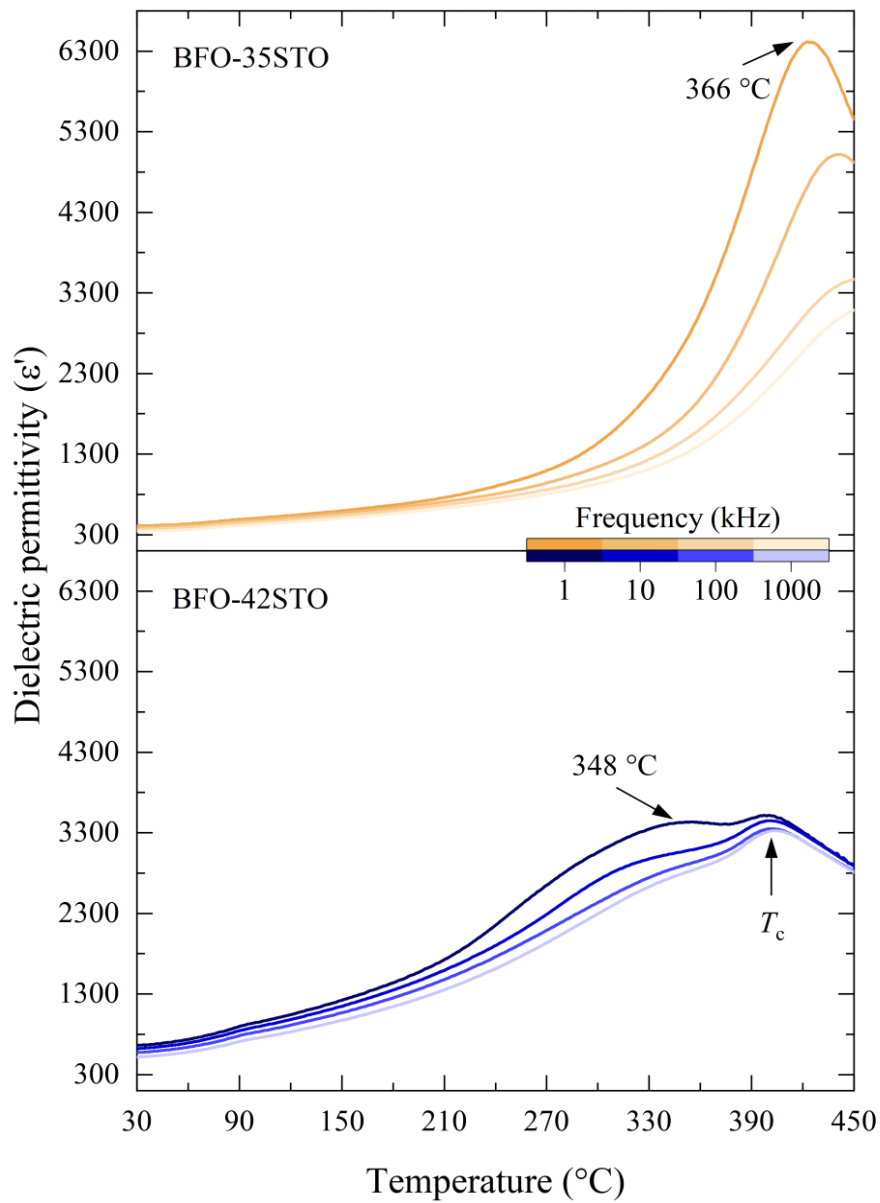

**Figure S2**. Temperature dependent dielectric permittivity for BFO-35STO and BFO-42STO at selected frequencies. The frequency dependent peak is characteristic of the relaxor behavior. The arrows at 1 kHz illustrate the peak shift towards lower temperatures, indicative of the increased ergodicity in BFO-42STO sample. Increasing the STO fraction the Curie temperature ($T_C$) is shifted to 401 °C, as seen by the frequency independent peak in BFO-42STO.